\documentclass[aps,prl,reprint, superscriptaddress]{revtex4-1}
\usepackage{bm}
\usepackage[dvips]{graphicx}
\usepackage{amsmath,amssymb}
\usepackage{epsfig}
\usepackage{color}
\usepackage{dcolumn}
\usepackage{comment}

\newcommand{\ketg}{\mbox{$|g\rangle $}}
\newcommand{\ketk}{\mbox{$|k\rangle $}}
\newcommand{\brag}{\mbox{$\langle g|$}}
\newcommand{\brak}{\mbox{$\langle k|$}}

\begin{document}
\title{Ultrafast Quantum-path Interferometry Revealing the Generation Process of Coherent Phonons} 

\author{Kazutaka G. Nakamura}
\email[Corresponding author: ]{nakamura@msl.titech.ac.jp}
\affiliation{Laboratory for Materials and Structures, Institute of Innovative Research, Tokyo Institute of Technology, 4259 Nagatsuta, Yokohama 226-8503, Japan}

\author{Kensuke Yokota}
\affiliation{Laboratory for Materials and Structures, Institute of Innovative Research, Tokyo Institute of Technology, 4259 Nagatsuta, Yokohama 226-8503, Japan}

\author{Yuki Okuda}
\affiliation{Laboratory for Materials and Structures, Institute of Innovative Research, Tokyo Institute of Technology, 4259 Nagatsuta, Yokohama 226-8503, Japan}

\author{Rintaro Kase}
\affiliation{Laboratory for Materials and Structures, Institute of Innovative Research, Tokyo Institute of Technology, 4259 Nagatsuta, Yokohama 226-8503, Japan}

\author{Takashi Kitashima}
\affiliation{Laboratory for Materials and Structures, Institute of Innovative Research, Tokyo Institute of Technology, 4259 Nagatsuta, Yokohama 226-8503, Japan}

\author{Yu Mishima}
\affiliation{Laboratory for Materials and Structures, Institute of Innovative Research, Tokyo Institute of Technology, 4259 Nagatsuta, Yokohama 226-8503, Japan}

\author{Yutaka Shikano}
\affiliation{Quantum Computing Center, Keio University, 3-14-1 Gakuen-cho, Yokohama, 223-8522, Japan}
\affiliation{Research Center for Advanced Science and Technology (RCAST), The University of Tokyo, 4-6-1 Komaba, Meguro, Tokyo 153-8904, Japan}
\affiliation{Research Center of Integrative Molecular Systems (CIMoS), Institute for Molecular Science, National Institutes of Natural Sciences, 38 Nishigo-Naka, Myodaiji, Okazaki, Aichi 444-8585, Japan}
\affiliation{Institute for Quantum Studies, Chapman University, 1 University Dr., Orange, California 92866, USA}

\author{Yosuke Kayanuma}
\email{kayanuma.y.aa@m.titech.ac.jp}
\affiliation{Laboratory for Materials and Structures, Institute of Innovative Research, Tokyo Institute of Technology, 4259 Nagatsuta, Yokohama 226-8503, Japan}
\affiliation{Graduate School of Sciences, Osaka Prefecture University, 1-1 Gakuen-cho, Sakai, Osaka, 599-8531 Japan}

\date{\today}

\begin{abstract}
Optical dual-pulse pumping actively creates quantum-mechanical superposition of the electronic and phononic states in a bulk solid.
We here made transient reflectivity measurements in an n-GaAs using a pair of relative-phase-locked femtosecond pulses and found characteristic interference fringes.
This is a result of quantum-path interference peculiar to the dual-pulse excitation as indicated by theoretical calculation.
Our observation reveals that the pathway of coherent phonon generation
in the n-GaAs is impulsive stimulated Raman scattering at the displaced potential due to the surface-charge field,
even though the photon energy lies in the opaque region.
\end{abstract}

\maketitle
Coherent control is a technique of manipulating quantum states in materials using optical pulses \cite{Brumer1986, Rice1986, Scherer1991}. 
A wave packet in quantum mechanical superposition is created by the optical pulse via several quantum transition paths.  
In the case of a double-pulse excitation, wave packets created by transitions in each pulse and across the two pulses interfere and the generated superposition state is manipulated by controlling a delay between the two pulses \cite{Katsuki2018, Mashiko2018}.  
A contribution of individual quantum paths can be extracted from the interference pattern, which is referred to as quantum-path interferometry \cite{Austin2011}.

\par
Coherent phonons are a temporally coherent oscillation of the optical phonons induced by the impulsive excitation of an ultrashort optical pulse \cite{Nelson1985,Merlin1997,Dekorsy2000, Misochko2001, Randi2017, Glerean2018}.
Using coherent phonons and making a pump--probe-type optical measurement, we can directly observe the dynamics of the electron--phonon coupled states in the time domain for a wide variety of materials \cite{Zeiger1992, DeCamp2001, Katsuki2013, Cho1990, Dekorsy1993, Merlin1996, Misochko2000, Hase2003, Wu2008, 
Kamaraju2010, Norimatsu2014, Misochko2015, Riffe2007, Sun2017}. 
In this respect, the clarification of the generation mechanism of the coherent phonon is a fundamental subject as an ultrafast dynamical process \cite{Pfeifer1992,Kuznetsov1995,Stevens2002}.
The generation mechanisms of coherent phonons are usually categorized as two types: a mechanism of impulsive stimulated Raman scattering (ISRS) \cite{Nelson1985} and a mechanism of displaced enhanced coherent phonons \cite{Zeiger1992}. In addition, for polar semiconductors such as GaAs, the screening of the surface-space-charge field \cite{Cho1990, Pfeifer1992} is considered to be another generation mechanism for opaque conditions. 
The generation mechanism of coherent phonons may become a controversial subject 
in the case of opaque-region pumping because impulsive absorption (IA) and ISRS processes coexist as possible quantum mechanical transition paths \cite{Nakamura2015}.  
A novel experimental technique is needed to shed light on this subject.

\par
In the present work, we apply quantum-path interferometry to study the generation process of coherent optical phonons through the coherent control of electron-phonon coupled states in bulk solids.
Coherent phonons are often coherently controlled using a pair of femtosecond pulses as pump pulses, and the phonon amplitude is enhanced or suppressed via the constructive or destructive interference of induced phonons \cite{Hase1996, Katsuki2013}.
Unlike these earlier works, we used two relative-phase-locked pump pulses (pulses 1 and 2) and a delayed probe pulse (pulse 3) \cite{Hayashi2014}, for quantum-path interferometry.  
If the delay of the dual-pump pulses $t_{12}$ was controlled with subfemtosecond accuracy and if the electronic coherence was maintained during the dual pulses, electronic excited states were created as a quantum mechanical superposition;  i.e., the electronic polarizations induced by pulses 1 and 2 interfered with each other. 
Meanwhile, polarization in the phonon system was coherently created with resulting interference within the phononic and electronic degrees of freedom. 
The probe pulse was used to monitor the interference fringe via heterodyne detection; i.e., via a change in the reflectivity as a function of the pump--pump delay  $t_{12}$ and pump--probe delay $t_{13}$. 
We could evaluate the electronic and phononic coherence times of the sample using this scheme.
Furthermore, a theoretical estimation predicts a decisive difference in the interference fringes between ISRS and IA, and it will be shown that the dominant pathway of the generation of coherent phonons can be determined from the pump-pulse-delay-dependent interference pattern of generation efficiency. 
\begin{figure}[htb]
\includegraphics[width=7.0cm]{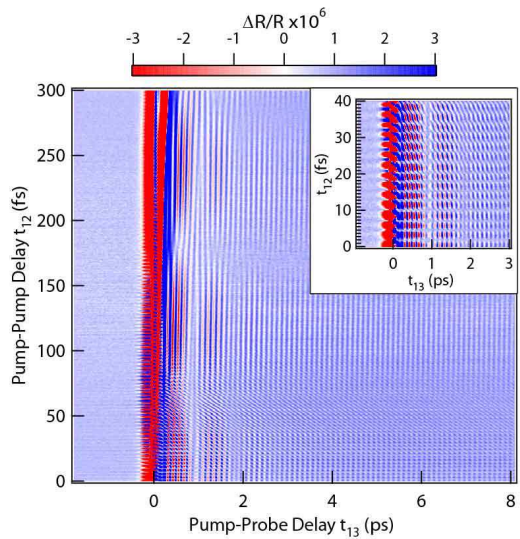}
\caption{Two-dimensional image map of the change in reflection intensity with the pump--probe delay ($t_{13}$) and pump--pump delay ($t_{12}$).}
\label{IIDmap}
\end{figure}

A femtosecond optical pulse (central wavelength of 798 nm, pulse width of $\sim$50 fs) was split with a partial beam splitter into two pulses (i.e., pump and probe pulses). The pump pulse was introduced into a homemade Michelson-type interferometer to produce relative-phase-locked pump pulses (i.e., pulses 1 and 2), in which stability was within 6 \%. 
The probe pulse (i.e., pulse 3) was irradiated with a controlled time delay.
The optical bandpass filter (the center wavelength of 800 nm with the band width of 10 nm) was used for detecting the reflected probe pulse in order to reduce cancellation effects of Stokes and anti-Stokes components~\cite{Nakamura2016}.
We set the two pump pulses in a collinear condition with parallel polarization in the present experiments. The $k$-vector direction of the two interfering pump pulses may affects to the interference fringes.
The sample was a single crystal of n-GaAs with (100) orientation and kept at 90 K in a cryostat. 
Details of the experimental setting are described in the Supplemental Material \cite{Supplement}.

Figure 1 is a two-dimensional map of the transient reflectivity change $\Delta R/R$ 
plotted against 
the pump--probe delay $t_{13}$ and pump--pump delay $t_{12}$. For a fixed value of  $t_{12}$, 
 $\Delta R/R$ indicates an oscillation with periods of 115 and 128 fs, which are equal 
to the periods of the longitudinal-optical (LO) phonon and LO phonon-plasmon coupled mode (LOPC) at the $\Gamma$ point in GaAs \cite{Mooradian1967,Lee2008,Ishioka2011,Hu2012}. 
For a fixed value of $t_{13}$, meanwhile, $\Delta R/R$ has a beat between 
a rapid oscillation with a period of 2.7 fs, which is nearly equal to that of the pump laser with a wavelength of 798 nm, 
and a slow oscillation with vibration periods of the LO phonon and LOPC. 
\par
For fixed values of $t_{12}$, a Fourier transformation of the signal $\Delta R/R$ was carried out with respect to $t_{13}$. 
The Fourier transformation was performed over the interval of 
$0.25 \ {\rm ps} < t_{13}<2.25 \ {\rm ps}$ after the irradiation of pump 2 to avoid the spurious effect of excess charge in the very early stage and 
the effect of phonon decay at a late time.
 Figure 2 is a plot of the oscillation amplitude $\Delta R^0$ of the Fourier-transformed data against the frequency $\omega_{13}$ and delay $t_{12}$.  
The figure shows that two modes of coherent oscillations are excited 
in the crystal. 

\par
\begin{figure}[htb]
\centering
\includegraphics[width=7.0 cm]{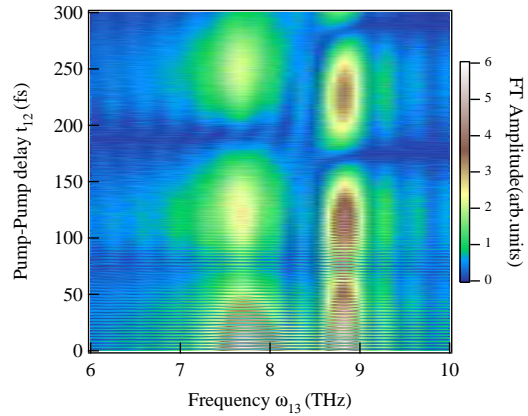}
\caption{Two-dimensional image map of the Fourier spectra at various pump--probe delays ($t_{13}$). }
\label{FTTmap}
\end{figure}

\par
To see more clearly the $t_{12}$ dependence of the oscillation amplitude $\Delta R^0$ 
of the transient reflectivity, 
we plotted $\Delta R^{0}$ at peak values at frequency $\omega_{13}$ of 8.7 and 7.8  ${\rm THz}$ for the LO and LOPC modes, respectively.
Figure 3 (a) presents results for the LO mode. The LOPC mode shown in Fig. 3(b) has qualitatively the same interference pattern as the LO mode. 
In Fig. 3 (a), the rapid oscillation with a period of $\sim 2.7$ fs is the interference fringe of the electronic 
states that memorized the phase of the pump-laser field. The slow oscillation with period $\sim 115$ fs is the interference fringe due to the coherence of phonons. 
The rapid interference fringes disappeared when we used the cross polarized pump pulses.
\par
Note that the electronic coherence survives well after the overlapping of the pump pulses ends, 
as can be seen from comparison with the linear optical interference of the dual pulses (Fig. 3 (c)). 
This means that the optical phase of pulse 1 is imprinted on the electronic polarization and interferes 
with that of pulse 2. 
The most important feature of the interference pattern is the apparent collapse and revival of the electronic fringe 
at around $t_{12}\sim 55$ fs. It will be shown below that this is due to a quantum-path interference 
peculiar to the ISRS process. 
\par
\begin{figure}[htb]
\centering
\includegraphics[width=7.0 cm]{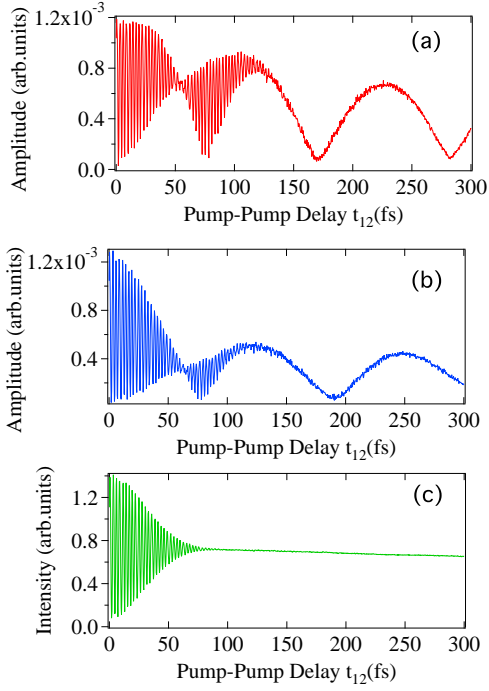}
\caption{ Interference fringe of LO phonon (a) and LOPC (b) and optical interference (c).}
\label{LOExp}
\end{figure}

In the time-region where $t_{13}$ is large enough compared with $t_{12}$ and the pulse-width, it is safely assumed that 
the generation process and the detection process of coherent phonons are well separated. 
Hereafter, we concentrate on the generation process of LO phonons. 
See Supplemental Material \cite{Supplement} for the theoretical treatment of the probe processes. 
For microscopic interactions that induce the coherent oscillation of LO phonons through the irradiation of 
ultrashort optical pulses, several models are conceivable, including the Fr\"ohlich interaction \cite{Frohlich1954}   
and deformation-potential interaction \cite{Shen1965}. 
In the case of polar materials, it is considered that electrostatic interaction due to 
transient depletion field screening plays a central role \cite{Pfeifer1992}. 
It is known that there are two types of photoinduced current, the usual injection current following the real excitation of carriers and the shift current resulting from quantum mechanical polarization induced by optical pulses  \cite{Kuznetsov1993, Sipe2000}, in ionic semiconductors \cite{Nastos2006}. 
The response of the shift current is usually faster than that of the injection current. 

\par
We assume a model Hamiltonian that describes the electron--phonon interaction as

\begin{eqnarray}
H&=&\left\{ \epsilon_g+\hbar\omega b^\dagger b\right\}\ketg \brag \nonumber\\
&+&\sum_k \left\{\epsilon_k+\hbar\omega b^\dagger b+\alpha \hbar\omega \left(b+b^\dagger\right)\right\}\ketk\brak,
\label{Hamiltonian}
\end{eqnarray}
where $\ketg$ is the electronic ground state of the crystal with energy $\epsilon_g$ and $\ketk$ the excited state 
with energy $\epsilon_k$. 
The creation and annihilation operators of the LO phonon at the $\Gamma$ point with energy $\hbar\omega$ are respectively denoted $b^\dagger$ and $b$. 
It is assumed that the dimensionless electron-phonon coupling constant $\alpha$ is  small and $k$-independent,  assuming a rigid-band shift.
The parameter $\alpha$ indicates the displacement of the potential, where all effects on deformation of the potential, such as the surface-space-charge field, are included.
 See the Supplemental Material \cite{Supplement} for a detailed explanation.  
\par 
Within the rotating-wave approximation, the interaction Hamiltonian with a dual-pump pulse is given by
\begin{equation}
H_{pump}(t)=E_{pu}(t)\sum_k\mu_k |k\rangle\langle g|+ H. c.,
\label{pump}
\end{equation}
where $\mu_k$ is the transition dipole moment from $\ketg$ to $\ketk$. $E_{pu}(t)$ is the temporal profile 
of the electric field of the pump pulse,
\begin{equation}
E_{pu}(t)=E_0\left( f(t)e^{-i\Omega_0 t}+f(t-t_{12})e^{-i\Omega_0(t-t_{12})}\right),
\label{Epump}
\end{equation}
where $\Omega_0$ is the carrier frequency of the laser pulse.
Here, $f(t)$ is the pulse envelope, which is assumed to have a Gaussian form,
$
f(t)=\left(1/\sqrt{\pi}\sigma\Omega_0\right) e^{-t^2/\sigma^2},
$
and $E_0$ is the amplitude of the electric field. 
A fundamental quantity used to describe the optical properties of crystals is the electric response function 
given by
\begin{equation}
F(t)=\sum_k|\mu_k|^2 e^{-i(\epsilon_k - \epsilon_g) t/\hbar -\eta |t|/\hbar},\quad (\eta=0_+),
\label{response}
\end{equation}
which is obtained via the Fourier transform of the effective optical absorption spectrum $I_{eff}\left(\Omega\right)$.

\par
\begin{figure}[htb]
\centering
\hspace{1cm}
\includegraphics[width=7.0 cm]{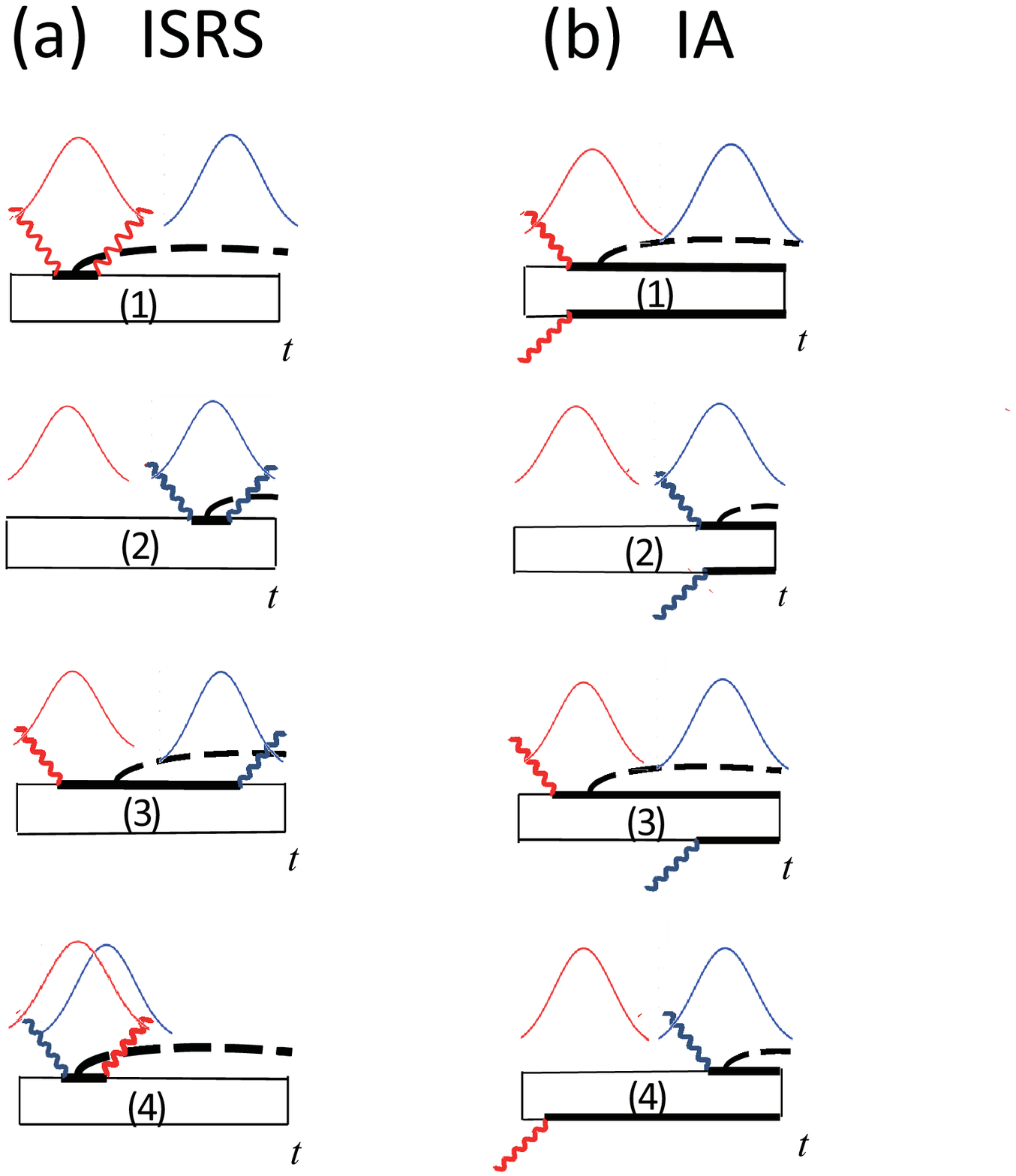}
\caption{Double-sided Feynman diagrams for the density matrices corresponding to (a) the ISRS process and (b) the IA process. The thin and thick solid lines respectively represent the ground and excited states. The dashed curves represent the one-LO-phonon state. 
The red and blue Gaussian curves represent the pulse envelope of the first and the second pulses, respectively, with the wavy lines their photon propagators.}
\label{Feynman3}
\end{figure}
\par
We adopt the density matrix formalism to derive the generation amplitude of the coherent phonon. 
The change of the amplitude $\Delta R$ of the reflectivity is proportional to the expectation value of the LO phonon coordinate $Q=\sqrt{\hbar/2\omega}\left ( b+b^\dagger \right )$ except for constant factors. See Supplemental Material \cite{Supplement} for the formula of spectrally resolved detection of  reflectivity modulation. 
Figure \ref{Feynman3} presents double-sided Feynman diagrams for the generation by ISRS (Fig. \ref{Feynman3} (a)) and IA (Fig. \ref{Feynman3} (b)). 
In Fig. \ref{Feynman3}, the propagators shown by thin lines correspond to the ground state and those shown by bold lines correspond to the excited state. The dashed lines represent the one-phonon state. 
Note that the Hermitian conjugate terms arise from the processes in the diagrams in which the upper and the lower propagators are interchanged, but these processes are ignored 
in Fig. \ref{Feynman3} (a) for simplicity. 

\par
After a perturbation calculation, the amplitude of the oscillation of coherent phonons in the ISRS and IA processes, 
$A _{ISRS}$ and $A_{IA}$ are respectively given as 
\begin{eqnarray}
A_{i}(t_{12})&=&
L_{i}(0)+e^{i\omega t_{12}}L_{i}(0) +e^{-i(\Omega_0-\frac{\omega}{2})t_{12}}L_{i}(t_{12}) \nonumber\\
&+& e^{i(\Omega_0+\frac{\omega}{2})t_{12}}L_{i}(-t_{12}), \label{A}
\end{eqnarray} 
in which $i=ISRS, IA$ and
\begin{eqnarray}
L_{ISRS}(x)&=&2i \int_0^\infty du \ g(u-x) \sin \frac{\omega u}{2}e^{i\Omega_0 u }F(u), \\
L_{IA}(x)&=&\int_{-\infty}^\infty du \ g (u-x)  e^{i(\Omega_0-\frac{\omega}{2}) u} F(u) ,
\end{eqnarray}
with $g(u) = e^{-u^2/(2 \sigma^2)}$, $x=0,t_{12},-t_{12}$. The amplitude $\Delta R^{(0)}$ is proportional to the absolute values of $A_{i}(t_{12})$. 
The first, second, third and fourth terms in Eq. (\ref{A}) correspond to the processes (1) to (4) in Fig. \ref{Feynman3}, respectively. 
Details of the calculation are shown in Supplemental Material \cite{Supplement}.

\par

\begin{figure}[htb]
\centering
\includegraphics[width=7.0cm]{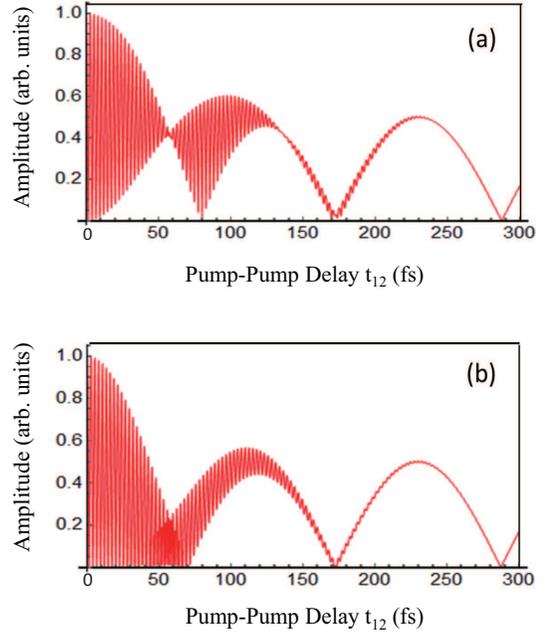}
\caption{(a) Theoretical curve for the interference fringe in the oscillation amplitude of transient reflectivity due to the ISRS process at the LO phonon frequency 
plotted against the 
pump--pump delay ($t_{12}$).
(b) Same as (a) but for the IA process.}
\label{LOTheory}
\end{figure}

The actual calculation of the transient reflectivity can be done for real materials if the electric 
response function $F(t)$ is given. 
In the calculation, we assumed a Lorentzian form,
$
I_{eff}(\Omega)=I_0\left( \Gamma/\pi\right)/\left\{(\Omega-\Omega_0)^2+\Gamma^2\right\},
$
with $\hbar\Omega_0=1.55$ eV and $\Gamma=0.015$ eV based on the absorption spectra \cite{Sturge1962, Casey1974}.
The calculated fringe patterns $\Delta R^{0}$ are shown for ISRS (Fig.\ref{LOTheory}(a)) and IA (Fig.\ref{LOTheory}(b)).  
We found that the features in the fringe shown in Fig. 3 (a) are well reproduced if it is assumed that only the ISRS process contributes to the generation of coherent phonons. Furthermore, the overall line shape is in good agreement with experimental data .

\par
Most important is the fact that the feature of the collapse and revival of the electronic fringe at around 
$t_{12} \sim 55$ fs arises only from the ISRS process, while the IA signal does not yield any such feature. 
This is due to the quantum-path interference peculiar to ISRS. In Fig. \ref{Feynman3} (a), the contribution arising from 
diagrams (1) and (2) gives rise only to the interference of the phonon, which is described by the first and second terms on the right-hand side of Eq. (\ref{A}). The electronic interference arises from diagram (3) and (4), which corresponds to the third and fourth terms respectively in Eq. (\ref{LOTheory}). 
It should be noted that the fourth term in $A_{ISRS}(t_{12})$ is negligibly small for $t_{12}>0$. 
Therefore, In ISRS, the electronic interference fringe appears only from the cross term between (1) + (2) and (3). At $t_{12}=\pi/\omega$, 
this term vanishes owing to the destructive interference of the phonon. The high-frequency oscillation of the electronic fringe therefore disappears at $t_{12}=\pi/\omega=55$ fs. 
This is a manifestation of the path interference of the electronic and phononic degrees of freedom in the dual-pump process peculiar to 
the ISRS. Note that in the IA process, both the third and fourth terms in $A_{IA}(t_{12})$ make a finite contribution so that the electronic fringe 
does not vanish at $t_{12}=\pi/\omega$.

\par
In Fig. \ref{LOTheory} (a), the amplitude of the electronic fringe becomes small for  $t_{12} > 130$ fs. 
This is due to the dephasing caused by the inhomogeneous broadening of the continuous spectrum in the excited states. 
In the experimental curve in Fig. \ref{LOExp}(a), the electronic fringe disappears almost completely for 
 $t_{12}>130$ fs in contrast to the case in Fig. \ref{LOTheory}. 
\par
The finding of the ISRS dominance in coherent phonon generation in the opaque region is surprising because, in the opaque region, the phonon generation intensity in the IA process is generally estimated to be higher 
than that for ISRS \cite{Stevens2002, Nakamura2015, Kayanuma2017}. 
We conjecture that even if the coherent phonon may be generated in the excited state subspace, 
its coherence is quickly lost because of the ultrafast deformation of the adiabatic potentials due to the electronic relaxation 
in the excited state of bulk materials. 
This may be one of the differences in the atomic and molecular dynamics of solids compared with those of the gas phase, 
in which the excited electronic states are long protected from relaxation. 
In addition, it was revealed that the generation of the coherent phonon in GaAs is a quick process as deduced from 
ISRS dominance even in the opaque region. The underlying mechanism is the quantum mechanically induced shift current. 
\par
In summary, we made transient reflectivity measurements for n-GaAs using relative-phase-locked femtosecond pulses and found characteristic interference fringes, which are assigned to quantum-path interference in the generation of coherent phonons.
Our observations and theory revealed that the pathway of coherent phonon generation in n-GaAs is ISRS at the displaced potential due to the surface-charge field, even though the photon energy lies in the opaque region. We demonstrated that optical dual-pulse pumping actively creates quantum-mechanical superposition of the electronic and phononic states in a bulk solid. 

\begin{acknowledgements}
The authors thank K. Norimatsu, K. Goto, H. Matsumoto, and F. Minami for their help with the experiments and calculation.
K. G. N., Y. S., and Y. K. thank K. Ohmori, H. Chiba, H. Katsuki, and Y. Okano of the Institute of Molecular Science for their valuable advice on the experiments.
This work was partially supported by Core Research for Evolutional Science and Technology of the Japan Science and Technology Agency, JSPS KAKENHI under grant numbers 25400330, 14J11318, 15K13377, 16K05396, 16K05410, 17K19051, and 17H02797, the Collaborative Research Project of Laboratory for Materials and Structures, the Joint Studies Program of the Institute of Molecular Science, National Institutes of Natural Sciences, and The Precise Measurement Technology Promotion Foundation. 
\end{acknowledgements}

\end{document}


\title{Supplementary Material: Ultrafast Quantum-path Interferometry Revealing the Generation Process of Coherent Phonons} 

\author{Kazutaka G. Nakamura}
\email[Corresponding author: ]{nakamura@msl.titech.ac.jp}
\affiliation{Laboratory for Materials and Structures, Institute of Innovative Research, Tokyo Institute of Technology, 4259 Nagatsuta, Yokohama 226-8503, Japan}

\author{Kensuke Yokota}
\affiliation{Laboratory for Materials and Structures, Institute of Innovative Research, Tokyo Institute of Technology, 4259 Nagatsuta, Yokohama 226-8503, Japan}

\author{Yuki Okuda}
\affiliation{Laboratory for Materials and Structures, Institute of Innovative Research, Tokyo Institute of Technology, 4259 Nagatsuta, Yokohama 226-8503, Japan}

\author{Rintaro Kase}
\affiliation{Laboratory for Materials and Structures, Institute of Innovative Research, Tokyo Institute of Technology, 4259 Nagatsuta, Yokohama 226-8503, Japan}

\author{Takashi Kitashima}
\affiliation{Laboratory for Materials and Structures, Institute of Innovative Research, Tokyo Institute of Technology, 4259 Nagatsuta, Yokohama 226-8503, Japan}

\author{Yu Mishima}
\affiliation{Laboratory for Materials and Structures, Institute of Innovative Research, Tokyo Institute of Technology, 4259 Nagatsuta, Yokohama 226-8503, Japan}

\author{Yutaka Shikano}
\affiliation{Quantum Computing Center, Keio University, 3-14-1 Gakuen-cho, Yokohama, 223-8522, Japan}
\affiliation{Research Center for Advanced Science and Technology (RCAST), The University of Tokyo, 4-6-1 Komaba, Meguro, Tokyo 153-8904, Japan}
\affiliation{Research Center of Integrative Molecular Systems (CIMoS), Institute for Molecular Science, National Institutes of Natural Sciences, 38 Nishigo-Naka, Myodaiji, Okazaki, Aichi 444-8585, Japan}
\affiliation{Institute for Quantum Studies, Chapman University, 1 University Dr., Orange, California 92866, USA}

\author{Yosuke Kayanuma}
\email{kayanuma.y.aa@m.titech.ac.jp}
\affiliation{Laboratory for Materials and Structures, Institute of Innovative Research, Tokyo Institute of Technology, 4259 Nagatsuta, Yokohama 226-8503, Japan}
\affiliation{Graduate School of Sciences, Osaka Prefecture University, 1-1 Gakuen-cho, Sakai, Osaka, 599-8531 Japan}

\date{\today}

\maketitle
\section{Experimental setup} \label{appendixA}
\begin{figure}[htb]
\centering
\includegraphics[width=8.5 cm]{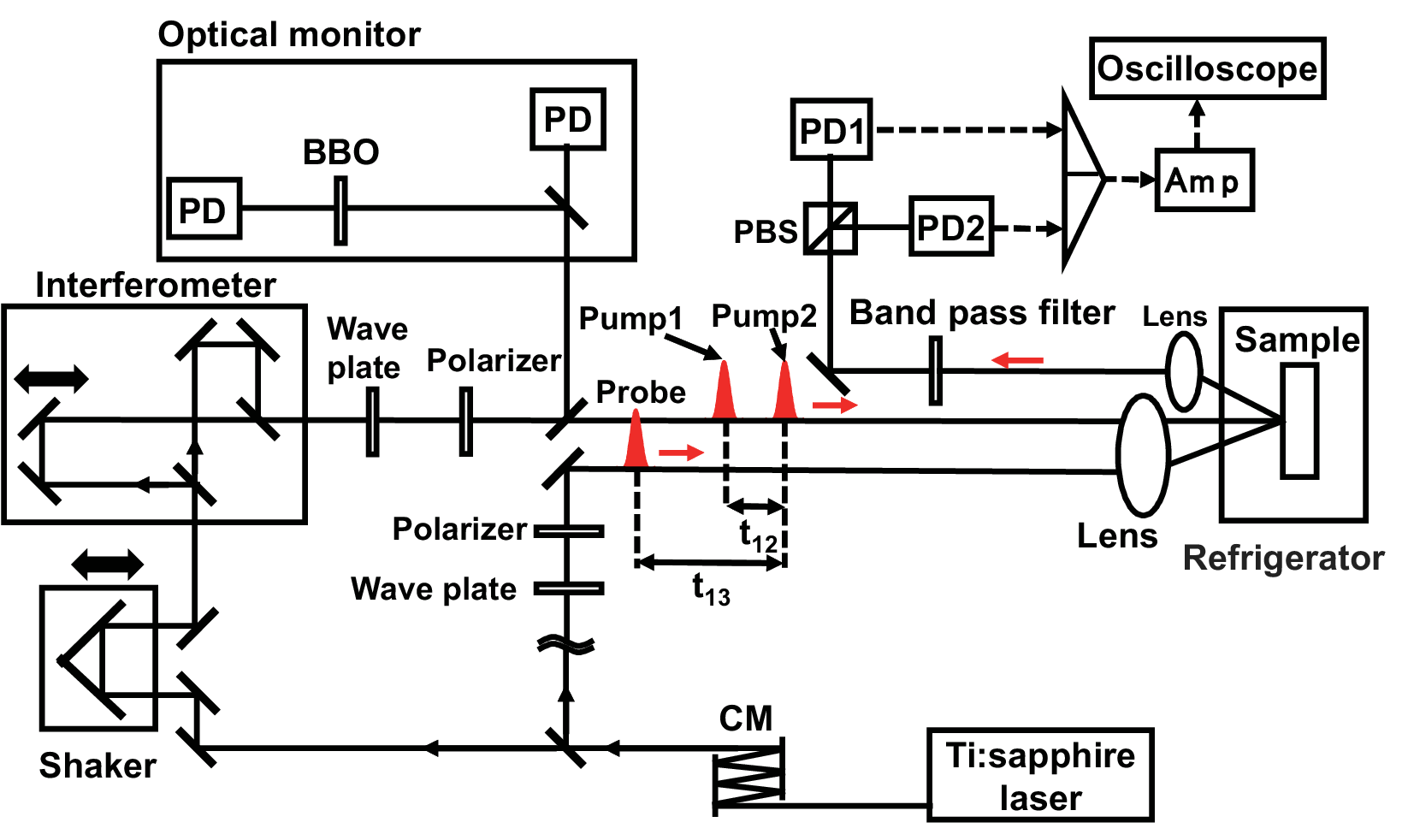}
\caption{Schematic of the experimental setup. CM, PD and BBO denotes chirped mirror, photodiode and BBO nonlinear optical crystal.}
\label{SetUp}
\end{figure}

\begin{figure}[htb]
\centering
\includegraphics[width=8 cm]{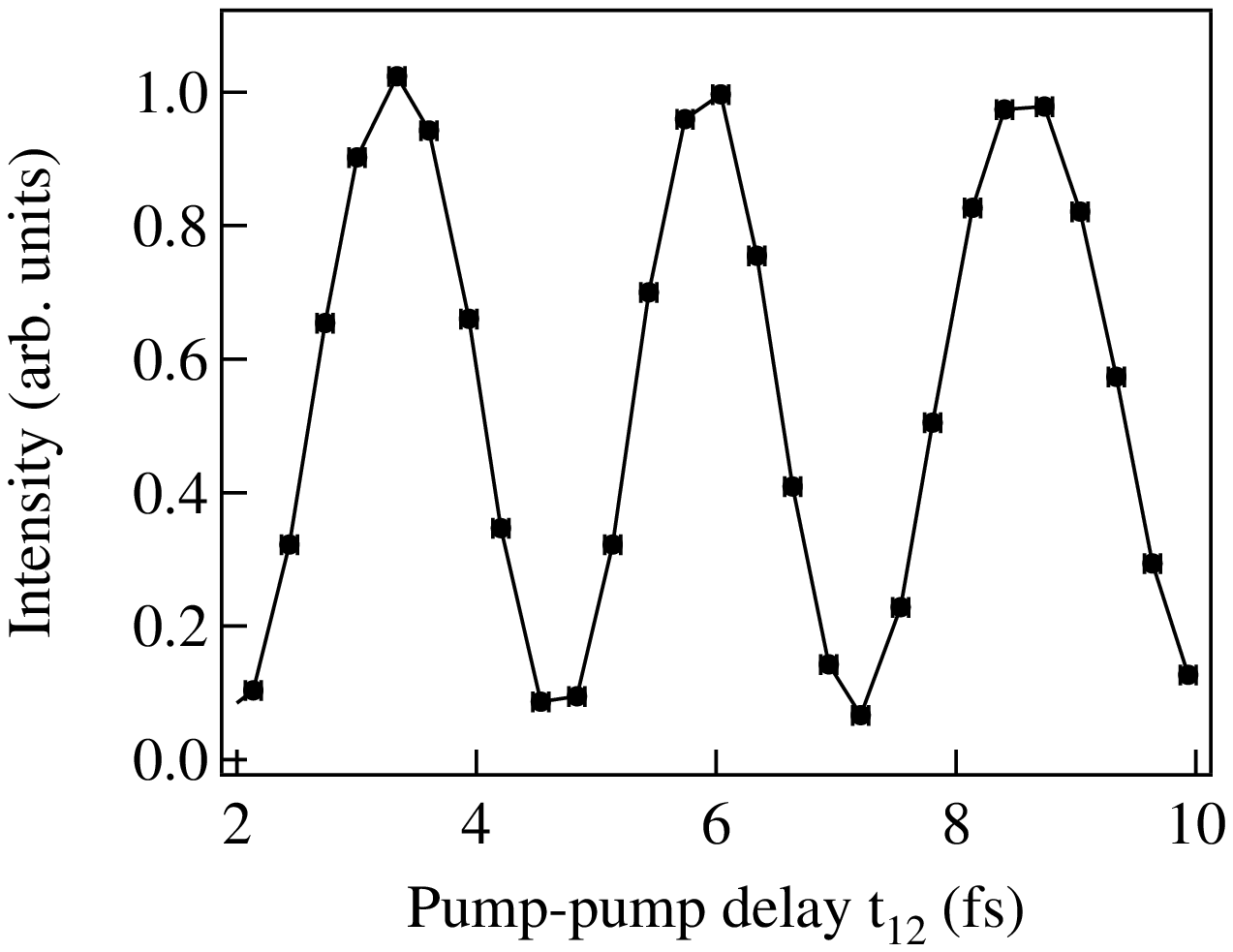}
\caption{Expanded figure of the optical interference between 2 and 10 fs presented with circles and lines.}
\label{Stage}
\end{figure}

A schematic of the pump-probe setup is shown in Fig. \ref{SetUp}. The output of the Ti:sapphire oscillator (center wavelength = 798 nm (1.54 eV), pulse width $\sim$ 50 fs) was split with a partial beam splitter into two pulses; these pulses were used as pump and probe pulses.    The pump pulse was introduces into a scan delay unit (APE Co.,  Scan Delay 150) to control the delay between the pump and probe pulses.  It was then introduced into a home-made Michelson-type interferometer to produce two relative phase-locked pump pulses (pulse 1 and 2).  One optical arm of the interferometer was equipped with an automatic positioning stage (Sigma Tech Co. Ltd., FS-1050SPX). 
The stage used is an active-feedback controlled stage (with a feedback stage controller FC-601A) with a minimum resolution of 5 nm and repetition-position accuracy of $\pm$10nm.  In the double-pulse experiments, we controlled the stage with a step of 45 nm, which corresponds to the optical-path change of 90 nm and the delay of approximately 0.3 fs.  The estimated phase stability is approximately 0.049 $\pi$ for the 800-nm light.  The expanded figure  of the optical interference between 2 and 10 fs presented with circles and lines in Fig. \ref{Stage}.

The power of pulses 1 and 2 were 18 and 21 mW, respectively. The power of the probe pulse was 8 mW.  The relative phase-locked pump pulses were focused on the sample using a lens, which also focused the probe pulse (pulse 3).   The sample was a single crystal of n-type GaAs with (100) orientation and set in a cryostat cooled by a refrigerator.  
The n-type GaAs was obtained from DOWA Semiconductor Co. (Si doped with  a carrier concentration of $9.6 \times 10^{19}$ cm$^{-3}$,  $350-{\rm \mu m}$  thickness and mirror-polished surface).
The refrigerator used is  Pascal-OP-S101 AD, which is the closed type GM cryocooler with a low-nose option in which the sample holder is apart from the cold head motor and attached to the laser table.  Temperature in the range between 10 and 300 K was controlled by using Model 9700 (Scientific Instruments inc.).
The pulse width was estimated from the linear optical interference and  the frequency resolved auto correlation (FRAC) for the pump pulses.  The pump and probe pulses were linearly polarized, and set along the (010) and ($\overline{1}$00) axis of GaAs, respectively.  The reflected pulse from the probe pulse was fed into a polarization beam splitter.  The parallel and perpendicular polarized light was monitored with two balanced photodiodes (PD1 and PD2).  The differential signal from the PDs was amplified with a current amplifier (SRS Co., SR570) and averaged in a digital oscilloscope (Iwatsu Co., DS-4354M).  The temporal evolution of the reflectivity change  ($\Delta R/R$)  was measured by scanning the delay ($t_{13}$) between the pump 1 and probe pulses, repetitively, at 20 Hz with a fixed $t_{12}$.  To filter non-oscillatory background in the temporal evolution, an electric bandpass filter (3-300 kHz) was used to amplify the oscillatory signals. The optical bandpass filter (the center wavelength of 800 nm with the band width of 10 nm) was used  for detecting the reflected probe pulse in order to reduce cancellation effects of Stokes and anti-Stokes components \cite{NNakamura2016}.

\section{Deformed Harmonic Potential} \label{appendixB}
The harmonic potential representing the optical phonon is deformed under a long range external field (e.g. the surface-charge field).  Here we treat a simple case for a uniform electric field ($F(q)=-dq$) acting along the phonon coordinate ($q$).  Although in realistic depletion layers the surface-charge field in nonuniform and decreases on the length scale of a few micrometers, the coherent carriers and phonon dynamics occur typically much shorter length scale.  Then the effective charge field can be treated as uniform in such a short scale.
When the external field potential is applied to a harmonic potential $U(q)= kq^2/2$, the potential changes to $U'(q)=U(q)+F(q)$;
\begin{equation}
U'(q) = \frac{k}{2}q^2-dq = \frac{k}{2} \left( q- \frac{d}{k} \right)^2 - \frac{d^2}{2k}.
\label{eq:1}
\end{equation}
Then the potential minimum position and energy shift $d/k$ and $-d^2/(2k)$, respectively. The slope $d$ of the external fields in the electronic excite state is lower than that in the ground state, because the surface screening is suppressed by a electron-hole pair or electronic polarization.  Then the effective harmonic potential in the excited state is displaced from that in the ground state.

\begin{figure}[htb]
\centering
\includegraphics[width= 7cm]{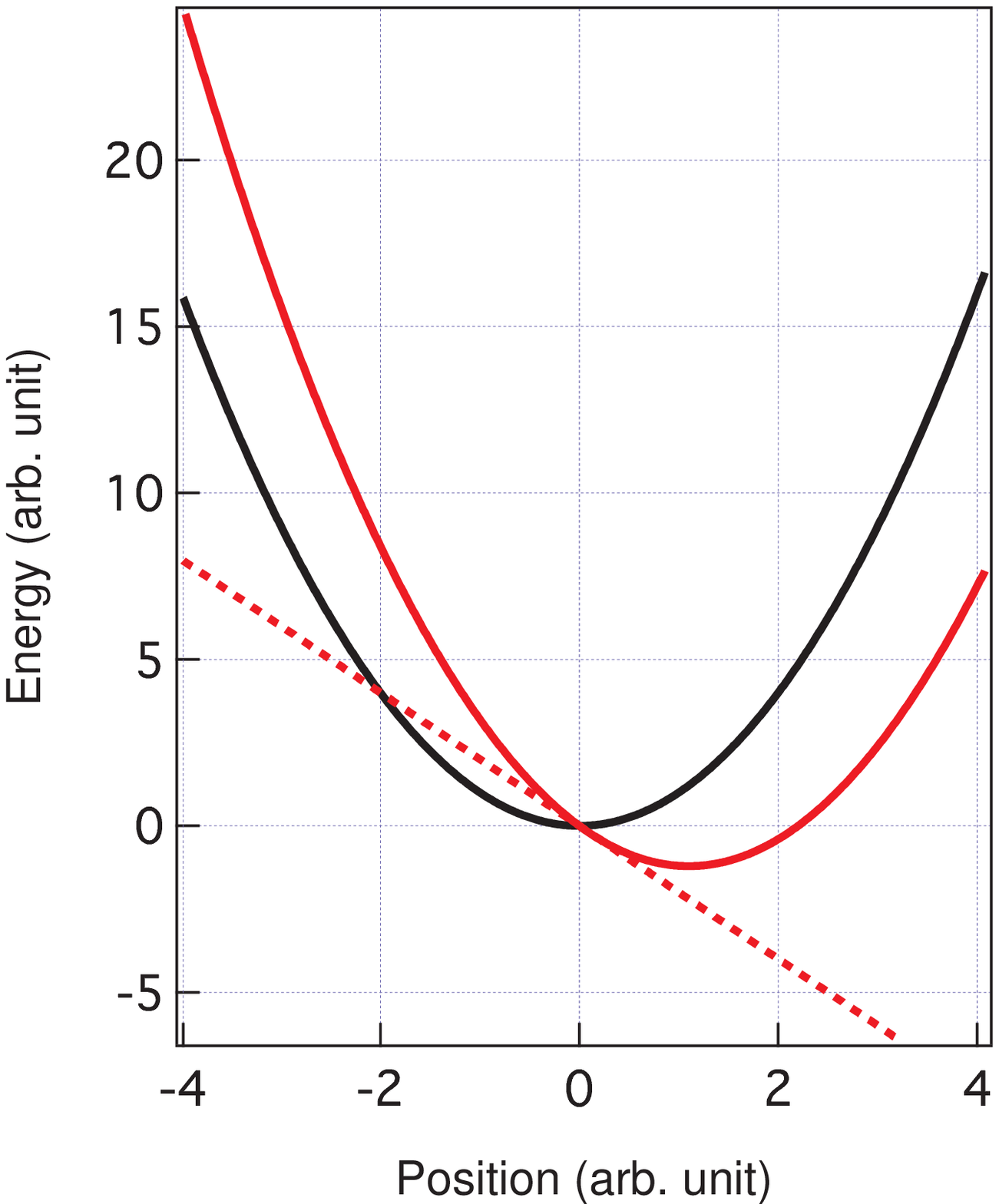}
\caption{The harmonic potential shifted by the external electric field potential. The solid black curve, dotted red line and solid red curve are the original harmonic potential ($U(q)$), external electronic field ($F(q)$) and the effective harmonic potential ($U'(q)$), respectively.}
\label{HO}
\end{figure}
\section{Calculation of Coherent Phonon Generation}\label{appendixC}
Here we describe details of the perturbation calculation of the density matrices for the coherent phonon generation by dual pulses, 
and see the origin of the difference in the interference fringes due to ISRS and IA processes. 
The equation of motion for the wave function $|\psi(t)\rangle$ in the Schr\"odinge picture,
\begin{equation}
 i\hbar\frac{d}{dt}| \psi(t)\rangle =\left\{H+H_{pump}(t)\right\}| \psi(t)\rangle 
\label{eq:2}
\end{equation}
 is transformed into that in the interaction picture by the 
substitution 
$|\tilde\psi(t)\rangle\equiv \exp[iH t/\hbar]|\psi(t)\rangle$ as 
\begin{equation}
 i\hbar\frac{d}{dt}|\tilde \psi(t)\rangle =\tilde H_{pump}(t)| \tilde\psi(t)\rangle , 
 \label{Dirac}
\end{equation} in which 
\begin{eqnarray}
&\tilde H&_{pump}(t)=e^{iHt/\hbar}H_{pump}(t)e^{-iHt/\hbar}\nonumber\\
&=&E_{pu}(t)\sum_k \mu_k \exp[i\left\{\epsilon_k+\hbar\omega b^\dagger b +\alpha\hbar \omega(b+b^\dagger)\right\}t/\hbar]\nonumber\\
&\times&\exp[-i\left\{\epsilon_g+\hbar\omega b^\dagger b \right\}t/\hbar]|k\rangle\langle g| + H.c. .
\label{eq:4}
\end{eqnarray}
Using a property of Boson operators, 
\begin{eqnarray} &\exp&[i\omega t\left\{b^\dagger b + \alpha (b + b^\dagger)\right\} ]\exp[-i\omega t b^\dagger b] \nonumber\\
&=&\exp[-i\alpha^2(\omega t - \sin \omega t)] \exp[g^*(t)b- g(t) b^\dagger ],
\label{eq:5}
\end{eqnarray}
with $g(t)=\alpha (1-e^{i\omega t})$, we find
\begin{eqnarray}
&\tilde H_{pump}& (t) = E_0\left\{f(t)e^{-i\Omega_0 t}+f(t-t_{12})e^{-i\Omega (t-t_{12})}\right\}\nonumber\\
&\times& \sum_k \mu_k \exp[i(\epsilon_k-\epsilon_g)t/\hbar-i\alpha^2(\omega t 
-\sin \omega t)]\nonumber\\
&\times &\exp[g^*(t) b-g(t)b^\dagger]|k\rangle\langle g|  + H.c. .
\label{eq:6}
\end{eqnarray}
\par
We calculate the density matrix to the second order with respect to $E_0$ and to the first order with respect to $\alpha$ 
under the condition that it is given by $|g,0\rangle \langle g,0|$ at $t=-\infty$, where  $|\xi,n\rangle$ denotes the ket vector 
for the $n$-phonon state $(n=0,1)$ in the ground state $(\xi=g)$ or the excited state $(\xi=e)$. 
For demonstration, the term corresponding to the paths (3) of ISRS shown in Fig. 4 (a) of the main text will be derived.  
The other terms are calculated by application of the same technique. 
In the ISRS paths, we need only to 
calculate the upper propagator (ket vector) since the system stays always in the ground states in the lower one.
In the path (3) of ISRS, the system is excited to $|e,0\rangle$ and $|e,1\rangle$ by the first pulse at time $t'$ and 
deexcited to $|g,1\rangle$ by the second pulse at $t''$. The formal solution to Eq. (\ref{Dirac}) written as 
\begin{equation}
 |\tilde \psi (t)\rangle =\exp_+[-\frac{i}{\hbar} \int_{-\infty}^t \tilde H_{pump} (\tau) d\tau ] |\tilde \psi (-\infty)\rangle 
 \label{eq:7}
 \end{equation}
is expanded to the second order terms. The term corresponding to the path (3) of ISRS is given by
\begin{eqnarray}
&|\tilde \psi (t)\rangle &=\alpha \Bigl(\frac{E_0}{\hbar} \Bigr)^2 \int_{-\infty}^t dt'' \int_{-\infty}^{t''} dt' f(t''-t_{12})f(t') \nonumber\\
&\times& F(t''-t') e^{i\Omega_0(t''-t'-t_{12})}(e^{i\omega t''}-e^{i\omega t'})|g,1\rangle .
\end{eqnarray}
The time-ordered integral can be reduced to a single integral by introducing a pair of new variables $(s,u)$ defined as 
$s=(t''+t')/2, u=(t''-t')/2$. If one notes $(t''-t_{12})^2+t'^2=2(s-t_{12})^2+\frac{1}{2}(u-t_{12})^2$, the integration over $s$ is 
carried out exactly and we find, except for constant factors, 
\begin{eqnarray}
|\tilde \psi (t)\rangle &= &2ie^{-i(\Omega_0-\frac{\omega}{2})t_{12}}\int_0^\infty du e^{-(u-t_{12})^2/(2\sigma^2)}\nonumber\\
&\times& \sin\frac{\omega u}{2} e^{i\Omega_0 u}F(u)|g,1\rangle. 
\end{eqnarray}
The amplitude of LO phonon oscillation is proportional to the expectation value 
$\langle g,0| Q |\tilde \psi (t)\rangle$. This reproduces the result in agreement with Eq. (6) in the main text.
\par
The process of path (4) in ISRS is \textit{anomalous}, since the system is excited by the second pulse and deexcited by the first pulse. 
Therefore, its contribution to the phonon generation is negligible except for the time region where the delay $t_{12}$ is very small. 
This is the reason why the apparent collapse and revival of the electronic fringe emerges in ISRS. Since, in the ISRS, the rapid frequency 
term $e^{-i\Omega_0 t_{12}}$ appears alone without counter-rotating term, the rapid oscillation is realized 
only through the cross term with 
the slowly oscillating terms due to paths (1) and (2). Thus it disappears for $t_{12}\simeq \pi/\omega$ at which the phonon oscillation 
becomes suppressed through the destructive interference of phonon. This is in sharp contrast to the IA process. In the IA, both of 
the path (3) and (4) have the same order of amplitude so that the electronic fringe does not disappear. The difference between the ISRS 
and IA originates from the difference in the integral domain over $u=(t''-t')/2$ as shown in Eqs. (6) and (7) in the main text. This is the manifestation of 
essential difference between the ISRS (light scattering) and IA (light absorption) in the generation mechanism of coherent phonons.

\section{Spectrally Resolved Detection of Reflectivity Modulation} \label{appendixD}
The interaction Hamiltonian for the probe pulse has the same form as Eq. (2) in the main text with the only change that $E_{pu}(t)$ is replaced by 
$
E_{pr}(t)=E_1f(t-t_{13}) e^{-i\Omega_0 t},
$
where $t_{13}$ is the delay time of the probe pulse and $E_1$ is its amplitude, which is usually much 
smaller than $E_0$. The coherent phonons generated by the pump pulses give rise to the periodic modulation $\Delta P(t)$ of the polarization 
due to the probe pulse. This results in the excess loss or gain of the reflected probe pulse. As shown in Ref. \cite{NNakamura2016},
the spectrally resolved reflection modulation $\Delta R(\Omega)$ for the detection frequency $\Omega$ and the delay $t_{13}$ is given by 
\begin{equation}
\Delta R(\Omega, t_{13})={\rm Im}\left\{E^*_{pr}(\Omega )\Delta P(\Omega)\right\},
\end{equation}
where $E_{pr}(\Omega)$ and $\Delta P(\Omega)$ are respectively the Fourier transforms of $E_{pr}(t)$ and $\Delta P(t)$. 
The spectrally resolved reflectivity $\Delta R(\Omega,t_{13})$ is a quantity of order of $|E_0|^2 |E_1|^2 \alpha^2$, and can be divided in 
two, $\Delta R(\Omega,t_{13})=\Delta R^{(a)}(\Omega,t_{13})+\Delta R^{(s)}(\Omega,t_{13})$, where $\Delta R^{(a)}(\Omega,t_{13})$ makes a main 
contribution to the anti-Stokes side, $\Omega\simeq \Omega_0 +\omega$, while $\Delta R^{(s)}(\Omega,t_{13})$ makes a main 
contribution to the Stokes side, 
$\Omega\simeq \Omega_0 -\omega$.
This is detected through the reflectivity change for the probe pulse 
that oscillates with the frequency of the LO phonon as a function of $t_{13}$.
\par
After a perturbation calculation, we find, aside from common constant factors, 
\begin{eqnarray}
&\Delta R_i^{(a)}&(\Omega,t_{13})=-\exp[-\frac{\sigma^2}{4}\{\left(\Omega-\Omega_0\right)^2+\left(\Omega-\Omega_0-\omega\right)^2\}] \nonumber\\
&\times& {\rm Re}\left\{A_i(t_{12}) e^{-i\omega t_{13}}
\left\{\chi (\Omega-\omega)-\chi(\omega)\right\}\right\}, \\
&\Delta R_i^{(s)}&(\Omega,t_{13})=-\exp[-\frac{\sigma^2}{4}\{\left(\Omega-\Omega_0\right)^2+\left(\Omega-\Omega_0+\omega\right)^2\}]  \nonumber\\
&\times& {\rm Re}\left\{A_i^*(t_{12}) e^{i\omega t_{13}}
\left\{\chi(\Omega+\omega)-\chi(\Omega)\right\}\right\},
\end{eqnarray}
where $\chi(\Omega)\equiv \int_0^\infty e^{i\Omega u }F(u) du$ is the optical susceptibility and $i=ISRS, IA$ correspond to the generation process ISRS and IA. 
In the above equations, $A_i(t_{12})$ is a quantity proportional to the expectation value of the amplitude of coherent phonon after the irradiation of the dual pulses, which are given in Eq. (5) in the main text. 
\par
As shown by the exponential factors, the above results indicate that the transient reflectivity oscillates with frequency $\omega$ both in the Stokes side and the anti Stokes 
side of the central frequency $\Omega_0$ of the probe pulse. The amplitude of the oscillation is proportional to the absolute values of 
$A_{ISRS}(t_{12})$ and $A_{IA}(t_{12})$. But it's relative phase is generally different depending on the complex amplitudes of $A_i(t_{12})$ and $\chi(\Omega)$. 
If $\chi(\Omega)$ is a smoothly varying function, we may set $\chi(\Omega\pm \omega)-\chi(\Omega)=\pm \frac{\partial \chi(\Omega)}{\partial \Omega}\omega$. Furthermore in the transparent region, $\chi(\Omega)$ is a real quantity. 
Then, the contribution from $\Delta R_i^{(a)}(\Omega,t_{13})$ and $\Delta R_i^{(s)}(\Omega,t_{13})$ almost cancels out in the spectrally integrated measurement \cite{NNakamura2016}.
This is approximately true even in the opaque region because of the cancellation due to the mismatching of the phase. This is the reason why the spectrally resolved measurement of transient reflectivity gives better quality of signals than the ordinary spectrally  integrated measurement.